\documentclass{INTERSPEECH2023}

\interspeechcameraready

\title{Exploration of Efficient End-to-End ASR using Discretized Input from Self-Supervised Learning}

\name{Xuankai Chang$^1$, Brian Yan$^1$, Yuya Fujita$^2$, Takashi Maekaku$^2$, Shinji Watanabe$^1$}
\address{
  $^1$Carnegie Mellon University, PA, USA\\
  $^2$Yahoo Japan Corporation, Japan}
\email{{xuankaic,byan,swatanab}@andrew.cmu.edu, {yuyfujit,tmaekaku}@yahoo-corp.jp}

\usepackage[
backend=biber,
style=ieee,
citestyle=numeric-comp,
maxbibnames=3,
maxcitenames=3,
doi=false,isbn=false,url=false,eprint=false
]{biblatex}

\defbibheading{bibliography}[\refname]{}

\DeclareSourcemap{
	\maps[datatype=bibtex, overwrite=true]{
		\map{
			\step[fieldsource=booktitle,
			match=\regexp{.*Interspeech.*},
			replace={Proc. Interspeech}]
			\step[fieldsource=journal,
			match=\regexp{.*INTERSPEECH.*},
			replace={Proc. Interspeech}]
			\step[fieldsource=booktitle,
			match=\regexp{.*ICASSP.*},
			replace={Proc. ICASSP}]
			\step[fieldsource=booktitle,
			match=\regexp{.*icassp_inpress.*},
			replace={Proc. ICASSP (in press)}]
			\step[fieldsource=booktitle,
			match=\regexp{.*Acoustics,.*Speech.*and.*Signal.*Processing.*},
			replace={Proc. ICASSP}]
			\step[fieldsource=booktitle,
			match=\regexp{.*International.*Conference.*on.*Learning.*Representations.*},
			replace={Proc. ICLR}]
			\step[fieldsource=booktitle,
			match=\regexp{.*International.*Conference.*on.*Computational.*Linguistics.*},
			replace={Proc. COLING}]
			\step[fieldsource=booktitle,
			match=\regexp{.*SIGdial.*Meeting.*on.*Discourse.*and.*Dialogue.*},
			replace={Proc. SIGDIAL}]
			\step[fieldsource=booktitle,
			match=\regexp{.*International.*Conference.*on.*Machine.*Learning.*},
			replace={Proc. ICML}]
			\step[fieldsource=booktitle,
			match=\regexp{.*North.*American.*Chapter.*of.*the.*Association.*for.*Computational.*Linguistics:.*Human.*Language.*Technologies.*},
			replace={Proc. NAACL}]
			\step[fieldsource=booktitle,
			match=\regexp{.*Empirical.*Methods.*in.*Natural.*Language.*Processing.*},
			replace={Proc. EMNLP}]
			\step[fieldsource=booktitle,
			match=\regexp{.*Association.*for.*Computational.*Linguistics.*},
			replace={Proc. ACL}]
			\step[fieldsource=booktitle,
			match=\regexp{.*Automatic.*Speech.*Recognition.*and.*Understanding.*},
			replace={Proc. ASRU}]
			\step[fieldsource=booktitle,
			match=\regexp{.*Spoken.*Language.*Technology.*},
			replace={Proc. SLT}]
			\step[fieldsource=booktitle,
			match=\regexp{.*Speech.*Synthesis.*Workshop.*},
			replace={Proc. SSW}]
			\step[fieldsource=booktitle,
			match=\regexp{.*workshop.*on.*speech.*synthesis.*},
			replace={Proc. SSW}]
			\step[fieldsource=booktitle,
			match=\regexp{.*Advances.*in.*neural.*information.*processing.*},
			replace={Proc. NeurIPS}]
			\step[fieldsource=booktitle,
			match=\regexp{.*Advances.*in.*Neural.*Information.*Processing.*},
			replace={Proc. NeurIPS}]
			\step[fieldsource=booktitle,
			match=\regexp{.*Workshop.*on.* Applications.* of.* Signal.*Processing.*to.*Audio.*and.*Acoustics.*},
			replace={Proc. WASPAA}]
			\step[fieldsource=publisher,
			match=\regexp{.+},
			replace={{}}]
			\step[fieldsource=month,
			match=\regexp{.+},
			replace={{}}]
			\step[fieldsource=location,
			match=\regexp{.+},
			replace={{}}]
			\step[fieldsource=address,
			match=\regexp{.+},
			replace={{}}]
			\step[fieldsource=organization,
			match=\regexp{.+},
			replace={{}}]
		}
	}
}
\begin{document}

\maketitle
 
\begin{abstract}
Self-supervised learning (SSL) of speech has shown impressive results in speech-related tasks, particularly in automatic speech recognition (ASR). 
While most methods employ the output of  intermediate layers of the SSL model as real-valued features for downstream tasks, there is potential in exploring alternative approaches that use discretized token sequences. 
This approach offers benefits such as lower storage requirements and the ability to apply techniques from natural language processing. 
In this paper, we propose a new protocol that utilizes discretized token sequences in ASR tasks, which includes de-duplication and subword modeling to enhance the input sequence. 
It reduces computational cost by decreasing the length of the sequence. 
Our experiments on the LibriSpeech dataset demonstrate that our proposed protocol performs competitively with conventional ASR systems using continuous input features, while reducing computational and storage costs.

\end{abstract}

\section{Introduction}
\label{sec:intro}

Over the past decade, remarkable advancements have been made in automatic speech recognition (ASR), largely due to the rapid development of deep neural networks~\cite{graves2006connectionist,abdel2012applying,graves2012sequence,graves2013speech,chan2015listen,vaswani2017attention,dong2018speech,gulati2020conformer}. These networks have significantly expanded the capabilities of speech recognition models. Additionally, the increasing availability of computing resources has enabled the training of ASR models using vast amounts of transcribed data, resulting in further improved performance ~\cite{panayotov2015librispeech,chen2021gigaspeech}. However, since deep neural networks require substantial amounts of data, some researchers have sought to increase their capacity by incorporating more transcribed data~\cite{chan2021speechstew}. Nevertheless, this approach has limitations, as a significant portion of available data remains untranscribed. To address this, researchers have proposed leveraging untranscribed data through unsupervised and semi-supervised learning techniques~\cite{lee2013pseudo,synnaeve2019end,kahn2020self}. Among these methods, self-supervised learning (SSL)~\cite{baevski2020wav2vec,hsu2021hubert,chen2022wavlm,baevski2022data2vec,mohamed2022self} has achieved impressive results in speech related downstream tasks~\cite{yang2021superb,}. There are several methods to make models more suitable for various downstream tasks, including fine-tuning pre-trained models~\cite{baevski2020wav2vec,hsu2021hubert}, extracting robust speech features~\cite{chang2021exploration}, and inserting adapters~\cite{thomas2022efficient}.

Advances in speech processing technology have led to the development of various applications that improve the convenience of human life, such as voice enabled robots and smart speaker. These advancements have greatly enhanced the ability of machines to interact with humans. However, the collection of speech data through such systems raises concerns about privacy~\cite{nautsch2019preserving}. Users may worry about the potential for their personal information to be leaked during data transmission or due to security issues in the storage system of such systems. One solution to these concerns is to transform speech signals into a different form of encoding that does not contain speaker-specific information while retaining the essential linguistic information.

In a study by Van et al. \cite{van2017neural}, it was observed that training a vector quantization-variational autoencoder (VQ-VAE) model with general speech representation learning can extract speaker-independent features from speech. Using VQ-VAE, the speech is discretized and encoded in discrete tokens, where each token represents the speech information in a short time interval. Later, several other SSL-based methods have been developed for learning general speech representations~\cite{baevskivq,baevski2020wav2vec,hsu2021hubert,chen2022wavlm}. These SSL models can also be used to discretize speech either through the vector quantization module in the model \cite{baevskivq, baevski2020wav2vec}, or by applying k-means clustering on hidden embeddings from these models \cite{hsu2021hubert, chen2022wavlm}. Note that it was shown that the higher layers of SSL models retain less speaker information than the lower layers, and the discretization step can further cleanse speaker-specific information.
Using discretized tokens from an SSL model as speech representations has several other advantages:
\begin{enumerate}
    \item Small storage and transmission size.
    \item Preservation of original speech duration information.
    \item Intermediate representation with both acoustic and linguistic information but with less speaker specific information.
\end{enumerate}
Previous studies have investigated the use of discrete-token input for ASR models~\cite{baevskivq,baevski2020effectiveness}, where 13.5 thousand unique discrete tokens are used. However, this approach did not outperform conventional log-mel-filterbank features in terms of ASR performance, unless a small BERT \cite{kenton2019bert} model trained with discretized token sequences is additionally used in front of ASR model.
As a result, the wide application of discrete token input in speech-processing tasks may be limited.

Given the advantages of using discretized token and the result of previous study, we are motivated to investigate using discrete tokens extracted from state-of-the-art (SOTA) SSL models to replace conventional speech processing inputs such as raw wave or acoustic features, such as mel filter bank~\cite{davis1980comparison}. In our study, we use WavLM~\cite{chen2022wavlm} to extract speech representations and a k-means model to obtain discrete tokens. Specifically, maximum $2,000$ unique tokens are used to represent the speech features, less than that in previous study~\cite{baevski2020effectiveness}. Once the speech data is discretized, it can be used in training and inference.
The discrete tokens offer a significant reduction in data size. For example, in certain conditions, 1,000 hours of speech data can be compressed from around 100 GB to less than 1 GB, which can be conveniently loaded to RAM at once. To prove the feasibility of this approach, we conduct experiments in end-to-end (E2E) speech recognition (ASR) task, using sequence-to-sequence (Seq2Seq) models. So far, we only reduce the size of data in terms of storage. However, it doesn't affect the computation efficiency, which is dominated by the model size and sequence length. To address the latter factor, we propose two methods. First, we can combine consecutively repeated tokens into a single one~\cite{lee2022direct}. Second, we apply subword modeling on discrete tokens~\cite{hayashi2020discretalk} to divide the sequence into subsequences, each of which is represented as a single meta-token. We evaluated our method on LibriSpeehch 100 hr subset for fast evaluation and LibriSpeech 960 hr. All pre-trained models and source code required for conducting experiments will be made publicly available with a license that allows free usage for research purposes.

\section{Speech Recognition Using Discretized Representation}
\label{sec:discrete_asr}

\begin{figure}
    \centering
    \includegraphics[width=\linewidth]{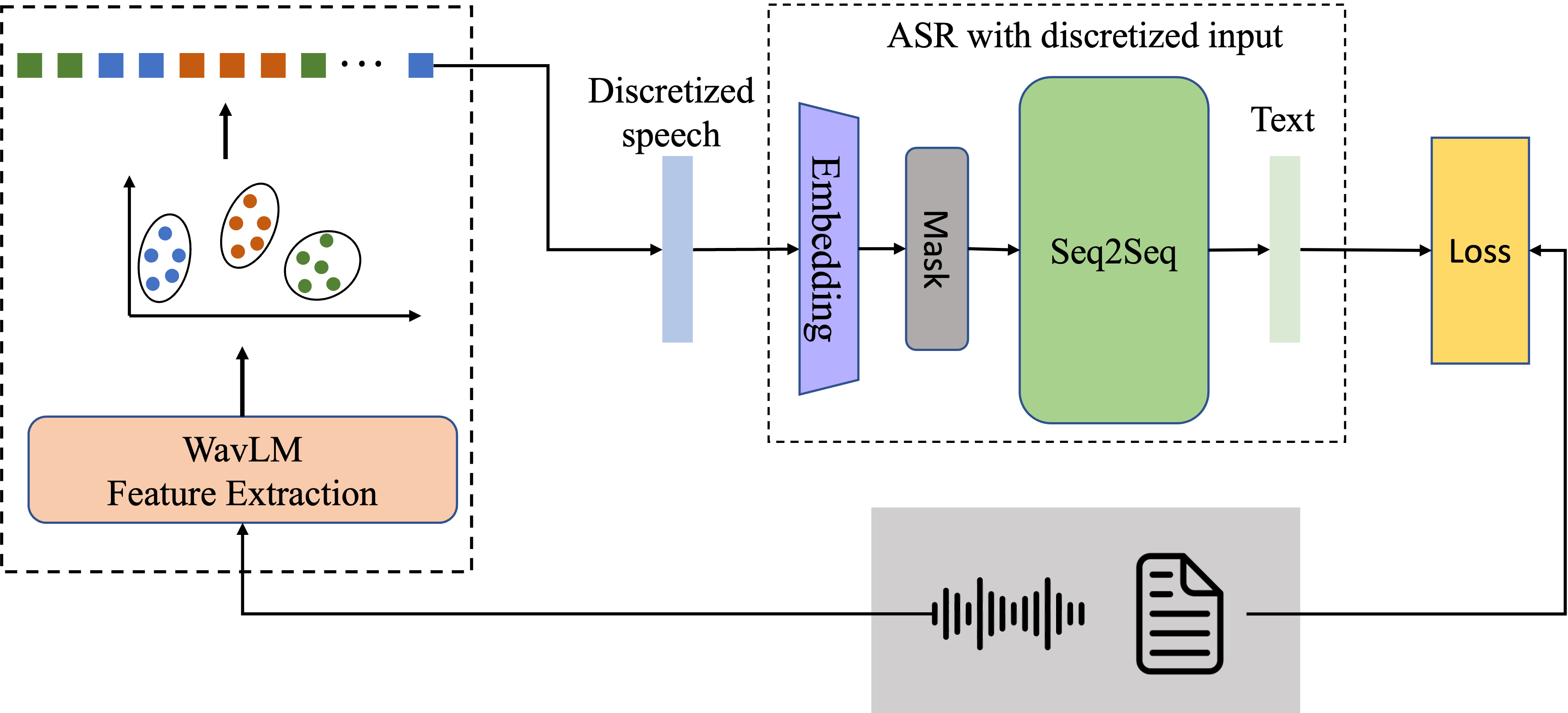}
    \caption{Illustration of proposed ASR with discretized input. The speech discretization process is shown on the left. On the right side is Seq2Seq model from discrete tokens to text (upper). }
    \label{fig:model}
\end{figure}

In this section, we present the details of the speech recognition model using discrete tokens in details, shown in Fig.~\ref{fig:model}, including data processing, speech recognition models and data augmentation methods. 

\subsection{Discretized Speech and Length Reduction}
\label{ssec:speech_disc_len_red}

This section provides a detailed description of the speech discretization process. Instead of using the original continuous speech or acoustic features, discrete tokens are employed as input in our proposed ASR model. This approach is similar to the one used in pre-training HuBERT~\cite{hsu2021hubert}. k-means clustering is applied to the hidden embeddings from intermediate SSL models to generate discretized speech encoding. It is worth noting that different features can be used for k-means clustering, depending on the task, including various SSL models or pre-trained supervised-learning models. Additionally, other discretization techniques, such as those proposed in~\cite{van2017neural,zeghidour2021soundstream,wang2023neural,defossez2022high}, can also be used.

In our experiments, we use a pre-trained WavLM model\footnote{We use WavLM Large model from \url{https://github.com/microsoft/unilm/tree/master/wavlm}}~\cite{chen2022wavlm} to extract hidden embeddings. It was trained primarily on learning the masked language models by predicting the pseudo-labels while performing speech denoising simultaneously. It achieved the best results on many downstream tasks in the SUPERB~\cite{yang2021superb}. Note that the WavLM large model used comprises 24 hidden layers. The last Transformer-Encoder layer from the model is chosen to extract hidden embeddings because it has been reported that the last layer contributes the most to ASR downstream tasks~\cite{masuyama2022end,chen2022wavlm}, according to a learned weight vector used to combine the hidden embeddings from all layers during training the ASR model. A k-means is trained to cluster the hidden embeddings and obtain the cluster indices as discrete tokens used as input to the ASR model. The resulting discretized input sequence has the same length as the sequence of hidden embeddings from the WavLM model, which produces speech features at a rate of 50 frames per second.

In Tab.\ref{tab:data_size}, we compare the data sizes of different data formats including raw waveform, conventional acoustic features, SSL-based features, and discrete tokens. Let us take a single-channel speech utterance of $T$ seconds as an example. The size of raw waveform data depends on the sampling rate and the audio sample encoding. Here we take the common settings used in speech recognition, i.e. 16 kHz wav in 16-bit signed integer form. For acoustic features, we use a $D$-dimensional float (4 Bytes or 32 bits) vector with a frame shift of $10$ms, which corresponds to a rate of 100 frames per second. An example value of $D$ is 80 used in \cite{guo2021recent}. We illustrate the SSL-based features using a single-layer hidden embedding from WavLM model stored as float vectors. While for the discrete tokens, we take the WavLM features clustered with a k-means model of maximum 4096 clusters (i.e. 12-bit).
\begin{table}[htbp]
    \centering
    \caption{Data size (bit) comparison of a T-second utterance among: 1) 16 kHz raw waveform in 16-bit wav format; 2) acoustic features in $D$-dimensional float vector with 100 frames/sec.; 3) SSL-based features using WavLM; 4) discrete tokens in 12-bit with 50 frames/sec.}
    \begin{tabular}{l|r}
    \toprule
    Data format    & Data size (bits) \\
    \midrule
    Raw waveform    & $ 16 \times 16000  \times T$ \\
    Acoustic features & $ 32 \times D \times 100 \times T $ \\
    SSL-based features & $ 32 \times 1024 \times 50 \times T $ \\
    Discrete tokens & $ 12 \times 50 \times T $ \\
    \bottomrule
    \end{tabular}
    \label{tab:data_size}
\end{table}

While this rate is efficient in most cases, we can further improve computational efficiency by \textbf{removing repeated tokens} and applying \textbf{subword modeling}. In~\cite{lee2022direct}, consecutively repeated discrete tokens are combined into a single token. Subword modeling, originally proposed in~\cite{sennrich-etal-2016-neural,kudo2018subword} to address the open vocabulary problem in text, was applied to discrete speech tokens from vector-quantization~\cite{hayashi2020discretalk} to reduce time resolution by identifying and grouping frequently occurring patterns as a single meta-token. To achieve this, we use Sentencepiece\footnote{\url{https://github.com/google/sentencepiece}} with unigram model~\cite{kudo2018subword}.

\subsection{ASR with Discretized Input}
\label{ssec:disc_asr_model}

This section provides an overview of the model utilized in our study. First, every $D$-dimensional continuous SSL feature vector of speech data is mapped to an $N$-bit discrete token, which serves as an intermediate representation between the acoustic features and the linguistic units. For example, $D$ is 1024 and $N$ is 12 in our experiments. The model acts as a ``translator'' to convert discrete tokens into text transcriptions. The input sequence comprises of discretized speech, while the output sequence is the transcription text. Typically, the input sequence is longer than the output sequence. Monotonic alignment is maintained between the input and output, allowing us to incorporate the connectionist temporal classification (CTC)~\cite{graves2006connectionist} loss. To accomplish ASR with discretized input, we utilize the joint CTC/attention-based encoder-decoder models that are used in acoustic feature-based end-to-end speech recognition systems~\cite{kim2017joint}. Note that a randomly initialized linear embedding layer is used before the encoder to extract learnable features for input discrete tokens.

\subsection{Data Augmentation for ASR with Discretized Input}
\label{ssec:data_aug}

In this section, we present the data augmentation method employed in our experiments.

\noindent\textbf{Time-masking} Masking is a fundamental technique that has been widely adopted in various machine learning tasks, such as speech recognition~\cite{park2019specaugment}, computer vision~\cite{devries2017improved}, and natural language processing~\cite{kenton2019bert}. This technique involves concealing a portion of the input data during the training phase, which forces the model to learn to make predictions based on incomplete information. Time masking is a specific type of masking where a continuous segment in the input sequence is masked, and the model is expected to predict the correct output based on the remaining information. 
By exposing the model to partially masked inputs, time-masking helps to improve the model's generalization ability and makes it more robust to different types of noise and signal perturbations. In our work, we apply multiple time masks on the embedding sequence of discretized tokens.

\section{Experimental Results}
\label{sec:exp}

\subsection{Setup}
\label{ssec:exp_setup}

In this study, we examined the effectiveness of the proposed protocol on the LibriSpeech corpus~\cite{panayotov2015librispeech}, which is a widely used benchmark dataset for ASR. To expedite the process of tuning hyper-parameters, we employed the subset comprised of 100 hour of clean speech as the primary dataset for conducting our experiments. All evaluations are performed on dev-\{clean,other\} and test-\{clean,other\} sets. When doing experiments with all 960 hours LibriSpeech data, speed perturbation with factors $0.9$ and $1.1$ are used to increase training data by two folds. However, no speed perturbation is applied for experiments with train\_clean\_100. All k-means models are trained using around 100 hours of speech from training sets. To this end, a 100-hour subset is randomly selected when using all LibriSpeech as the training set.

Our implementation is based on ESPnet~\cite{watanabe2018espnet}, an open-source toolkit for end-to-end speech processing. Our ASR models use the joint CTC/attention-based encoder-decoder architecture based on the E-Branchformer~\cite{kim2022branchformer}. The encoder consists of 12 blocks, each with 4 self-attention heads, a convolutional gating multi-layer perceptron (cgMLP), and a feed-forward network (FFN) with an intermediate hidden dimension of 1024. The cgMLP convolution kernel size is 31. For the decoder, we used a 6-layer Transformer with a FFN dimension of 2048. We set the dropout rate to 0.1 and used 5,000 BPE subword units for output tokens. We set the CTC weight to 0.3 and did not employ language models in our experiments. For ASR decoding, we set the CTC weight to 0.3 and the beam size to 20.

\subsection{LibriSpeech100 Baselines}
\begin{table}[htbp]
    \centering
    \caption{LibriSpeech-100 baseline WERs (\%) of continuous feature-based ASR model using log-Mel-Filterbank (FBank), WavLM-Large with weighted-sum of all layers' and WavLM-Large only last layer's acoustic features.}
    \scalebox{0.9}{
    \begin{tabular}{c|cc|cc}
        \toprule
        \multirow{2}{*}{Feature} & \multicolumn{2}{c}{dev} & \multicolumn{2}{|c}{test} \\
         & clean & other & clean & other \\
        \midrule
        FBank & 8.1 & 21.6 & 8.3 & 22.2 \\
        \midrule
        WavLM-Large weighted-sum & 3.5 & \textbf{6.0} & 3.5 & \textbf{6.2} \\
        \midrule
        WavLM-Large last-layer & \textbf{3.4} & 6.4 & \textbf{3.4} & 6.5 \\
        \bottomrule
    \end{tabular}
    \label{tab:ls100_baseline}}
\end{table}
The performance of the baseline systems is presented in Table~\ref{tab:ls100_baseline}. We employ three baseline systems to evaluate the performance of our proposed method. The first system uses log-MelFilterbank (FBank) features with a frame shift of 10ms, while the other two systems adopt self-supervised learning features previously used in~\cite{chang2021exploration}. The difference between the latter two is whether the acoustic model uses hidden representations from the last layer of the WavLM large model or a weighted-sum of all layers as input features. A convolutional subsampling layer is applied between the features and acoustic models to reduce the time resolution as the default setting, resulting in a frame shift of 40ms. SpecAugment~\cite{park2019specaugment} is applied to both FBank and WavLM speech features. The results show that using features from the WavLM-Large model can achieve significantly better performance, especially on the dev-other and test-other sets compared to FBank. Note that using only the embedding of the last-layer is slightly inferior to using a weighted-sum of all layers.

\subsection{Number of Discrete Tokens}
\label{ssec:num_kmeans_clusters}

\begin{table}[htbp]
    \centering
    \caption{LibriSpeech-100 WERs (\%) of discretized input-based ASR with various number of discrete tokens: \{100, 500, 1000, 2000\}. Phoneme purity (phn\_pur), discrete token purity (dsc\_pur) and phone-normalized mutual information (PNMI) are listed for each type of discrete tokens.}
    \scalebox{0.75}{
    \begin{tabular}{c|ccc|cc|cc}
        \toprule
        \multirow{2}{*}{\# of tokens} & \multicolumn{3}{c|}{k-means quality} & \multicolumn{2}{c}{dev} & \multicolumn{2}{|c}{test} \\
         & phn\_pur & dsc\_pur & PNMI & clean & other & clean & other \\
        \midrule
        100 & 0.5750 & 0.3183 & 0.5591 & 8.1 & 18.2 & 8.4 & 19.0  \\
        500 & 0.6732 & 0.1096 & 0.6740 & 6.8 & 14.5 & 6.8 & 14.9  \\
        1000 & 0.7075 & 0.0721 & 0.7075 & 5.8 & 13.0 & 6.2 & 13.2 \\
        2000 & 0.7357 & 0.0391 & \textbf{0.7394} & \textbf{5.6} & \textbf{12.5} & \textbf{5.9} & \textbf{12.8} \\
        \bottomrule
    \end{tabular}}
    \label{tab:ls100_kmeans_results}
\end{table}
We report the ASR performance of models trained on train\_clean\_100 dataset using discretized inputs with different discrete tokens numbers, as shown in Table~\ref{tab:ls100_kmeans_results}. Our experiments involve setting varying numbers of clusters for the k-means model to obtain discrete tokens. These tokens are fed into the seq2seq model as discretized inputs, without reducing the length. To evaluate the quality of k-means labels, we follow the approach of~\cite{hsu2021hubert} to compute the phoneme purity, discrete token purity, and phone-normalized mutual information (PNMI). These metrics are computed by comparing the discrete tokens with the phoneme alignment\footnote{It is based on a Montreal Forced Aligner, available at \url{https://zenodo.org/record/2619474\#.ZAblHuyZNqs}}~\cite{lugosch2019speech} using the dev sets. In~\cite{hsu2021hubert}, PNMI was used to assess the quality of discrete tokens. Our ASR results show that more discrete tokens lead to better PNMI, as well as better ASR performance. Using 2000 clusters yields the best word error rates (WERs) of $5.9\%$ and $12.8\%$ on test-clean and test-other, respectively. These two numbers are about $29\%$ and $42\%$ better than those of the FBank-based ASR system, however they are worse than the WavLM feature-based systems. 
Based on the results, we use 2000 as the number of tokens in the subsequent experiments.

\subsection{Length reduction method}
\label{ssec:len_red}

\begin{table}[htbp]
    \centering
    \caption{LibriSpeech-100 WERs (\%) of discretized input based ASR \ with different length reduction methods on 2000-token representation, including de-duplication (dedup), apply subword modeling (SW) and combined together. In \cite{baevski2020effectiveness}, a VQ-Wav2Vec is used to discretize speech (13.5K tokens), together with fine-tuning a pre-trained discrete token-BERT and using a 4-gram LM in decoding.}
    \scalebox{0.7}{
    \begin{tabular}{p{0.45\linewidth}|c|cc|cc}
        \toprule
        \multirow{2}{*}{Length Reduction} & Avg. Input Length & \multicolumn{2}{c|}{dev} & \multicolumn{2}{c}{test} \\
         & (train / dev) & clean & other & clean & other \\
        \midrule
        vq-wav2vec \cite{baevski2020effectiveness} + token-BERT + 4-gram LM  & - / - & 4.0 & 10.9 & 4.5 & 12.1 \\
        \midrule
        WavLM-token & 633.8 / 358.1 & 5.6 & 12.5 & 5.9 & 12.8 \\
        \hspace{0.3cm} +SW\_6000 & 349.9 / 200.1 & 5.5 & 11.8 & 5.8 & 12.0 \\
        \midrule
        WavLM-token dedup & 468.7 / 270.9 & 6.4 & 12.7 & 6.4 & 13.2 \\
        \hspace{0.3cm} +SW\_6000 & \textbf{249.4} / \textbf{144.5} & 5.8 & 12.3 & 6.0 & 12.5 \\
        \bottomrule
    \end{tabular}}
    \label{tab:ls100_km2000_tokenization_results}
\end{table}
We evaluated the ASR performance of models trained on the LibriSpeech-100 dataset using different methods to reduce the length of the input sequence. First, we used subword modeling alone and set the total number of subwords to 6,000 based on the original WavLM discrete tokens. This resulted in an average input sequence length reduction of approximately $44\%$, and improved the ASR performance by $6\%$ on test-other. Next, we applied de-duplication by combining consecutively repeated tokens from the original WavLM token sequence, which reduced the average input sequence length by about $24\%$. The ASR performance degrades by $3\%$ on test-other set. Finally, we applied subword modeling on the discrete token sequence after de-duplication, resulting in a total sequence length reduction of $60\%$. Using this shorter sequence, we achieved a final performance of $12.5\%$ on test-other. Among the above 4 different types of sequences, subword modeling on original discrete tokens achieves the best performance. Comparing our result against the results in the previous study~\cite{baevski2020effectiveness}, it is worse. Note that in~\cite{baevski2020effectiveness}, a VQ-Wav2Vec~\cite{baevskivq} with 13.5 thousand discrete tokens is used, together with a token-BERT fine-tuned on the 100-hour training set. A 4-gram language model (LM) was used in decoding. Given this, our method is much simpler while achieving close performance. Since applying both subword modeling and de-duplication achieves the best computation efficiency due to the shortest sequence length, we use them together in our subsequent experiments on the large-scale data for fast training speed.

\subsection{LibriSpeech960 Results}
\label{ssec:ls960}

\begin{table}[htbp]
    \centering
    \caption{WERs(\%) of continuous features and discretized input w/. and w/o. 1-D Convolutional downsampling at a rate of 2 (1D-Conv) on LibriSpeech960.}
    \scalebox{0.8}{
    \begin{tabular}{c|cc|cc}
        \toprule
        \multirow{2}{*}{Feature} & \multicolumn{2}{c|}{dev} & \multicolumn{2}{c}{test} \\
         & clean & other & clean & other \\
        \midrule
        ASR-FBank & 2.5 & 6.3 & 2.6 & 6.2 \\
        ASR-WavLM last-layer & 1.9 & 3.9 & 2.0 & 4.0 \\
        \midrule
        Discrete tokens (dedup) & 2.9 & 6.8 & 3.0 & 7.0 \\
        Discrete tokens (dedup+1D\_Conv) & 2.9 & 6.8 & 3.1 & 6.9 \\
        \bottomrule
    \end{tabular}}
    \label{tab:ls960_final_results}
\end{table}
In Table.~\ref{tab:ls960_final_results}, we present the performance on LibriSpeech 960 hours data with speed perturbation. As baselines, we listed the conventional FBank-based ASR system and the WavLM-Large feature-based ASR models. We can see that using discrete token input with 2000 unique tokens achieves slightly worse performance than the FBank. However, the gap is relatively small. More hyper-parameter searching efforts are required on LibriSpeech 960 hours, including the length reduction method, time masking ratio, number of k-means clusters and the vocabulary size of subword modeling, etc. We found that training ASR models using the discrete units on large-scale data can be quite efficient. We further reduce the input sequence length by employing a 1-D convolution layer with a downsampling rate of 2. It is observed that applying a downsampling layer didn't hurt the performance while reducing the length by $50\%$. About computation efficiency, it took us $23$ minutes/epoch to train a discrete token-based ASR model with 1-D convolution layer on 4 Nvidia V100 GPUs, which is about half of that using FBank features. We attributed this to three reasons. First, input sequence length was reduced significantly, resulting in small computation and memory footprint. Second, large batch sizes can be achieved given that the input sequences are short. Last, the I/O overhead can be reduced because all the training data can be loaded to RAM at the beginning. In theory, the total size of 960 hours of training data can be as small as 0.3 GB ($\approx 960 * 3600 * 50 * 12$ bits) \footnote{To clarify, we didn't implement the data storing part in bit in our implementation. Instead, we used the normal int32 for convenience and compatibility with neural network toolkits. However, in that case, the data is still less than 1 GB for 960 hours of speech.}.

\section{Conclusions}
\label{sec:conclusion}

In this paper, we proposed a new protocol for E2E-ASR that uses discrete tokens as input to replace conventional raw waveform or acoustic feature input. These tokens are computed as the k-means cluster indices of hidden embeddings derived from state-of-the-art semi-supervised learning (SSL) models, specifically WavLM. Using discrete speech data can considerably reduce the size of data required for transmission and storage. By implementing de-duplication and subword modeling, the sequence length can be reduced, resulting in better computation efficiency. We experimentally compared different combinations of length-reduction method and provide the results. In addition, the discrete tokens computed from hidden embeddings of SSL models trained in general representation learning may preserve less speaker information, thus providing the benefit of preserving speaker privacy. Compared with previous studies on ASR with discrete tokens, our proposed methods is more straightforward. The experiments on the LibriSpeech dataset show that the proposed methods can achieve competitive results compared to conventional acoustic features. Future research could involve investigating other discretization techniques and ensembling discrete tokens to further enhance speech recognition performance.


\section{Acknowledgements}
Some experiments of this work used the Bridges2 system at PSC and Delta system at NCSA through allocation CIS210014 from the Advanced Cyberinfrastructure Coordination Ecosystem: Services \& Support (ACCESS) program, which is supported by National Science Foundation grants \#2138259, \#2138286, \#2138307, \#2137603, and \#2138296. We also gratefully acknowledge the support of NVIDIA Corporation with the donation of the A6000 GPUs used for this research.



\section{References}
{
\printbibliography

@inproceedings{abdel2012applying,
  title={Applying convolutional neural networks concepts to hybrid {NN}-{HMM} model for speech recognition},
  author={Abdel-Hamid, Ossama and Mohamed, Abdel-rahman and Jiang, Hui and Penn, Gerald},
  booktitle={Proc. ICASSP},
  pages={4277--4280},
  year={2012},
  organization={IEEE}
}

@inproceedings{graves2013speech,
  title={Speech recognition with deep recurrent neural networks},
  author={Graves, Alex and Mohamed, Abdel-rahman and Hinton, Geoffrey},
  booktitle={Proc. ICASSP},
  pages={6645--6649},
  year={2013},
  organization={IEEE}
}

@inproceedings{chan2015listen,
  title={Listen, attend and spell},
  author={Chan, William and Jaitly, Navdeep and Le, Quoc V and Vinyals, Oriol},
  booktitle={Proc. ICASSP},
  pages={4960--4964},
  year={2016},
  organization={IEEE}
}

@inproceedings{vaswani2017attention,
  title={Attention is all you need},
  author={Vaswani, Ashish and Shazeer, Noam and Parmar, Niki and Uszkoreit, Jakob and Jones, Llion and others},
  booktitle={Proc. NeurIPS},
  pages={5998--6008},
  year={2017}
}

@inproceedings{dong2018speech,
  title={Speech-{T}ransformer: {A} no-recurrence sequence-to-sequence model for speech recognition},
  author={Dong, Linhao and Xu, Shuang and Xu, Bo},
  booktitle={Proc. ICASSP},
  pages={5884--5888},
  year={2018},
  organization={IEEE}
}

@inproceedings{gulati2020conformer,
  title={Conformer: {C}onvolution-augmented {T}ransformer for speech recognition},
  author={Gulati, Anmol and Qin, James and Chiu, Chung-Cheng and Parmar, Niki and others},
  booktitle={Proc. Interspeech},
  pages={5036--5040},
  year={2020},
  organization={ISCA}
}

@inproceedings{panayotov2015librispeech,
  title={Librispeech: {A}n {ASR} corpus based on public domain audio books},
  author={Panayotov, Vassil and Chen, Guoguo and Povey, Daniel and Khudanpur, Sanjeev},
  booktitle={Proc. ICASSP},
  pages={5206--5210},
  year={2015},
  organization={IEEE}
}

@inproceedings{chen2021gigaspeech,
  title={Giga{S}peech: {A}n evolving, multi-domain {ASR} corpus with 10,000 hours of transcribed audio},
  author={Chen, Guoguo and Chai, Shuzhou and Wang, Guanbo and Du, Jiayu and Zhang, Wei-Qiang and Weng, Chao and Su, Dan and Povey, Daniel and Trmal, Jan and Zhang, Junbo and others},
  booktitle={Proc. Interspeech},
  year={2021},
  organization={ISCA}
}

@inproceedings{chan2021speechstew,
  title={Speech{S}tew: {S}imply mix all available speech recognition data to train one large neural network},
  author={Chan, William and Park, Daniel and Lee, Chris and Zhang, Yu and Le, Quoc and Norouzi, Mohammad},
  booktitle={Proc. Interspeech},
  year={2021},
  organization={ISCA}
}

@article{baevski2020wav2vec,
  title={wav2vec 2.0: A framework for self-supervised learning of speech representations},
  author={Baevski, Alexei and Zhou, Yuhao and Mohamed, Abdelrahman and Auli, Michael},
  journal={Advances in neural information processing systems},
  volume={33},
  pages={12449--12460},
  year={2020}
}

@article{hsu2021hubert,
  title={Hubert: Self-supervised speech representation learning by masked prediction of hidden units},
  author={Hsu, Wei-Ning and Bolte, Benjamin and Tsai, Yao-Hung Hubert and Lakhotia, Kushal and Salakhutdinov, Ruslan and Mohamed, Abdelrahman},
  journal={IEEE/ACM Transactions on Audio, Speech, and Language Processing},
  volume={29},
  pages={3451--3460},
  year={2021},
  publisher={IEEE}
}

@inproceedings{chang2021exploration,
  title={An exploration of self-supervised pretrained representations for end-to-end speech recognition},
  author={Chang, Xuankai and Maekaku, Takashi and Guo, Pengcheng and Shi, Jing and Lu, Yen-Ju and Subramanian, Aswin Shanmugam and Wang, Tianzi and Yang, Shu-wen and Tsao, Yu and Lee, Hung-yi and others},
  booktitle={2021 IEEE Automatic Speech Recognition and Understanding Workshop (ASRU)},
  pages={228--235},
  year={2021},
  organization={IEEE}
}

@article{yang2021superb,
  title={Superb: Speech processing universal performance benchmark},
  author={Yang, Shu-wen and Chi, Po-Han and Chuang, Yung-Sung and Lai, Cheng-I Jeff and Lakhotia, Kushal and Lin, Yist Y and Liu, Andy T and Shi, Jiatong and Chang, Xuankai and Lin, Guan-Ting and others},
  journal={arXiv preprint arXiv:2105.01051},
  year={2021}
}

@inproceedings{lee2022direct,
  title={Direct Speech-to-Speech Translation With Discrete Units},
  author={Lee, Ann and Chen, Peng-Jen and Wang, Changhan and Gu, Jiatao and Popuri, Sravya and Ma, Xutai and Polyak, Adam and Adi, Yossi and He, Qing and Tang, Yun and others},
  booktitle={Proceedings of the 60th Annual Meeting of the Association for Computational Linguistics (Volume 1: Long Papers)},
  pages={3327--3339},
  year={2022}
}

@article{chen2022wavlm,
  title={Wavlm: Large-scale self-supervised pre-training for full stack speech processing},
  author={Chen, Sanyuan and Wang, Chengyi and Chen, Zhengyang and Wu, Yu and Liu, Shujie and Chen, Zhuo and Li, Jinyu and Kanda, Naoyuki and Yoshioka, Takuya and Xiao, Xiong and others},
  journal={IEEE Journal of Selected Topics in Signal Processing},
  volume={16},
  number={6},
  pages={1505--1518},
  year={2022},
  publisher={IEEE}
}

@article{hayashi2020discretalk,
  title={Discretalk: Text-to-speech as a machine translation problem},
  author={Hayashi, Tomoki and Watanabe, Shinji},
  journal={arXiv preprint arXiv:2005.05525},
  year={2020}
}

@inproceedings{baevskivq,
  title={vq-wav2vec: Self-Supervised Learning of Discrete Speech Representations},
  author={Baevski, Alexei and Schneider, Steffen and Auli, Michael},
  booktitle={International Conference on Learning Representations}
}

@article{masuyama2022end,
  title={End-to-End Integration of Speech Recognition, Dereverberation, Beamforming, and Self-Supervised Learning Representation},
  author={Masuyama, Yoshiki and Chang, Xuankai and Cornell, Samuele and Watanabe, Shinji and Ono, Nobutaka},
  journal={arXiv preprint arXiv:2210.10742},
  year={2022}
}

@article{van2017neural,
  title={Neural discrete representation learning},
  author={Van Den Oord, Aaron and Vinyals, Oriol and others},
  journal={Advances in neural information processing systems},
  volume={30},
  year={2017}
}

@article{defossez2022high,
  title={High fidelity neural audio compression},
  author={D{\'e}fossez, Alexandre and Copet, Jade and Synnaeve, Gabriel and Adi, Yossi},
  journal={arXiv preprint arXiv:2210.13438},
  year={2022}
}

@article{zeghidour2021soundstream,
  title={Soundstream: An end-to-end neural audio codec},
  author={Zeghidour, Neil and Luebs, Alejandro and Omran, Ahmed and Skoglund, Jan and Tagliasacchi, Marco},
  journal={IEEE/ACM Transactions on Audio, Speech, and Language Processing},
  volume={30},
  pages={495--507},
  year={2021},
  publisher={IEEE}
}

@article{wang2023neural,
  title={Neural Codec Language Models are Zero-Shot Text to Speech Synthesizers},
  author={Wang, Chengyi and Chen, Sanyuan and Wu, Yu and Zhang, Ziqiang and Zhou, Long and Liu, Shujie and Chen, Zhuo and Liu, Yanqing and Wang, Huaming and Li, Jinyu and others},
  journal={arXiv preprint arXiv:2301.02111},
  year={2023}
}

@inproceedings{sennrich-etal-2016-neural,
    title = "Neural Machine Translation of Rare Words with Subword Units",
    author = "Sennrich, Rico  and
      Haddow, Barry  and
      Birch, Alexandra",
    booktitle = "Proceedings of the 54th Annual Meeting of the Association for Computational Linguistics (Volume 1: Long Papers)",
    month = aug,
    year = "2016",
    address = "Berlin, Germany",
    publisher = "Association for Computational Linguistics",
    url = "https://aclanthology.org/P16-1162",
    doi = "10.18653/v1/P16-1162",
    pages = "1715--1725",
}

@inproceedings{kudo2018subword,
  title={Subword Regularization: Improving Neural Network Translation Models with Multiple Subword Candidates},
  author={Kudo, Taku},
  booktitle={Proceedings of the 56th Annual Meeting of the Association for Computational Linguistics (Volume 1: Long Papers)},
  pages={66--75},
  year={2018}
}

@inproceedings{kim2017joint,
  title={Joint CTC-attention based end-to-end speech recognition using multi-task learning},
  author={Kim, Suyoun and Hori, Takaaki and Watanabe, Shinji},
  booktitle={2017 IEEE international conference on acoustics, speech and signal processing (ICASSP)},
  pages={4835--4839},
  year={2017},
  organization={IEEE}
}

@inproceedings{graves2006connectionist,
  title={Connectionist temporal classification: labelling unsegmented sequence data with recurrent neural networks},
  author={Graves, Alex and Fern{\'a}ndez, Santiago and Gomez, Faustino and Schmidhuber, J{\"u}rgen},
  booktitle={Proceedings of the 23rd international conference on Machine learning},
  pages={369--376},
  year={2006}
}

@article{kim2022branchformer,
  title={E-Branchformer: Branchformer with Enhanced merging for speech recognition},
  author={Kim, Kwangyoun and Wu, Felix and Peng, Yifan and Pan, Jing and Sridhar, Prashant and Han, Kyu J and Watanabe, Shinji},
  journal={arXiv preprint arXiv:2210.00077},
  year={2022}
}

@inproceedings{watanabe2018espnet,
  author={Shinji Watanabe and Takaaki Hori and Shigeki Karita and Tomoki Hayashi and Jiro Nishitoba and Yuya Unno and Nelson {Enrique Yalta Soplin} and Jahn Heymann and Matthew Wiesner and Nanxin Chen and Adithya Renduchintala and Tsubasa Ochiai},
  title={{ESPnet}: End-to-End Speech Processing Toolkit},
  year={2018},
  booktitle={Proceedings of Interspeech},
  pages={2207--2211},
  doi={10.21437/Interspeech.2018-1456},
  url={http://dx.doi.org/10.21437/Interspeech.2018-1456}
}

@article{lugosch2019speech,
  title={Speech Model Pre-Training for End-to-End Spoken Language Understanding},
  author={Lugosch, Loren and Ravanelli, Mirco and Ignoto, Patrick and Tomar, Vikrant Singh and Bengio, Yoshua},
  journal={Proc. Interspeech 2019},
  pages={814--818},
  year={2019}
}

@inproceedings{lee2013pseudo,
  title={Pseudo-label: {T}he simple and efficient semi-supervised learning method for deep neural networks},
  author={Lee, Dong-Hyun and others},
  booktitle={Proc. ICML},
  pages={896},
  year={2013}
}

@inproceedings{synnaeve2019end,
  title={End-to-end {ASR}: {F}rom supervised to semi-supervised learning with modern architectures},
  author={Synnaeve, Gabriel and Xu, Qiantong and Kahn, Jacob and Likhomanenko, Tatiana and Grave, Edouard and Pratap, Vineel and Sriram, Anuroop and Liptchinsky, Vitaliy and Collobert, Ronan},
  booktitle={Proc. ICML},
  year={2020}
}

@inproceedings{kahn2020self,
  title={Self-training for end-to-end speech recognition},
  author={Kahn, Jacob and Lee, Ann and Hannun, Awni},
  booktitle={Proc. ICASSP},
  pages={7084--7088},
  year={2020},
  organization={IEEE}
}

@article{mohamed2022self,
  title={Self-supervised speech representation learning: A review},
  author={Mohamed, Abdelrahman and Lee, Hung-yi and Borgholt, Lasse and Havtorn, Jakob D and Edin, Joakim and Igel, Christian and Kirchhoff, Katrin and Li, Shang-Wen and Livescu, Karen and Maal{\o}e, Lars and others},
  journal={IEEE Journal of Selected Topics in Signal Processing},
  year={2022},
  publisher={IEEE}
}

@inproceedings{thomas2022efficient,
  title={Efficient adapter transfer of self-supervised speech models for automatic speech recognition},
  author={Thomas, Bethan and Kessler, Samuel and Karout, Salah},
  booktitle={ICASSP 2022-2022 IEEE International Conference on Acoustics, Speech and Signal Processing (ICASSP)},
  pages={7102--7106},
  year={2022},
  organization={IEEE}
}

@article{davis1980comparison,
  title={Comparison of parametric representations for monosyllabic word recognition in continuously spoken sentences},
  author={Davis, Steven and Mermelstein, Paul},
  journal={IEEE transactions on acoustics, speech, and signal processing},
  volume={28},
  number={4},
  pages={357--366},
  year={1980},
  publisher={IEEE}
}

@inproceedings{guo2021recent,
  title={Recent developments on espnet toolkit boosted by conformer},
  author={Guo, Pengcheng and Boyer, Florian and Chang, Xuankai and Hayashi, Tomoki and Higuchi, Yosuke and Inaguma, Hirofumi and Kamo, Naoyuki and Li, Chenda and Garcia-Romero, Daniel and Shi, Jiatong and others},
  booktitle={ICASSP 2021-2021 IEEE International Conference on Acoustics, Speech and Signal Processing (ICASSP)},
  pages={5874--5878},
  year={2021},
  organization={IEEE}
}

@article{devries2017improved,
  title={Improved regularization of convolutional neural networks with cutout},
  author={DeVries, Terrance and Taylor, Graham W},
  journal={arXiv preprint arXiv:1708.04552},
  year={2017}
}

@inproceedings{kenton2019bert,
  title={BERT: Pre-training of Deep Bidirectional Transformers for Language Understanding},
  author={Kenton, Jacob Devlin Ming-Wei Chang and Toutanova, Lee Kristina},
  booktitle={Proceedings of NAACL-HLT},
  pages={4171--4186},
  year={2019}
}

@article{park2019specaugment,
  title={SpecAugment: A Simple Data Augmentation Method for Automatic Speech Recognition},
  author={Park, Daniel S and Chan, William and Zhang, Yu and Chiu, Chung-Cheng and Zoph, Barret and Cubuk, Ekin D and Le, Quoc V},
  journal={Proc. Interspeech 2019},
  pages={2613--2617},
  year={2019}
}

@article{graves2012sequence,
  title={Sequence transduction with recurrent neural networks},
  author={Graves, Alex},
  journal={arXiv preprint arXiv:1211.3711},
  year={2012}
}

@inproceedings{baevski2022data2vec,
  title={Data2vec: A general framework for self-supervised learning in speech, vision and language},
  author={Baevski, Alexei and Hsu, Wei-Ning and Xu, Qiantong and Babu, Arun and Gu, Jiatao and Auli, Michael},
  booktitle={International Conference on Machine Learning},
  pages={1298--1312},
  year={2022},
  organization={PMLR}
}

@inproceedings{baevski2020effectiveness,
  title={Effectiveness of self-supervised pre-training for ASR},
  author={Baevski, Alexei and Mohamed, Abdelrahman},
  booktitle={ICASSP 2020-2020 IEEE International Conference on Acoustics, Speech and Signal Processing (ICASSP)},
  pages={7694--7698},
  year={2020},
  organization={IEEE}
}

@article{nautsch2019preserving,
  title={Preserving privacy in speaker and speech characterisation},
  author={Nautsch, Andreas and Jim{\'e}nez, Abelino and Treiber, Amos and Kolberg, Jascha and Jasserand, Catherine and Kindt, Els and Delgado, H{\'e}ctor and Todisco, Massimiliano and Hmani, Mohamed Amine and Mtibaa, Aymen and others},
  journal={Computer Speech \& Language},
  volume={58},
  pages={441--480},
  year={2019},
  publisher={Elsevier}
}
}

\end{document}